\documentclass{emulateapj}

\slugcomment{Accepted to AJ, June 2005}

\shorttitle{Gliese 337CD}
\shortauthors{Burgasser, Kirkpatrick \& Lowrance}

\begin{document}

\title{Multiplicity Amongst Wide Brown Dwarf Companions to Nearby Stars: Gliese 337CD}

\author{
Adam J.\ Burgasser\altaffilmark{1,2},
J.\ Davy Kirkpatrick\altaffilmark{3},
and Patrick J.\ Lowrance\altaffilmark{4}
}

\altaffiltext{1}{Department of Astrophysics,
Division of Physical Sciences,
American Museum of Natural History, Central Park West at 79$^{th}$ Street,
New York, NY 10024, USA; adam@amnh.org}
\altaffiltext{2}{Spitzer Fellow}
\altaffiltext{3}{Infrared Processing and Analysis Center, M/S 100-22,
California Institute of Technology, Pasadena, CA 91125; davy@ipac.caltech.edu}
\altaffiltext{4}{Spitzer Science Center, M/S 220-6.
California Institute of Technology, Pasadena, CA 91125; lowrance@ipac.caltech.edu}

\begin{abstract}
We present Lick Natural Guide Star Adaptive Optics observations of the L8
brown dwarf
Gliese 337C, which is resolved for the first time into two closely separated
(0$\farcs$53$\pm$0$\farcs$03), nearly equal magnitude components with a $K_s$
flux ratio of 0.93$\pm$0.10.  Companionship is inferred from the
absence of a 3$\farcs$6 offset source in 2MASS or photographic plate images, implying
that the observed secondary component is a co-moving late-type dwarf.
With a projected separation of 11 AU and nearly equal-magnitude components, Gliese 337CD
has properties similar to other known companion and field
substellar binaries.  Its long orbital period (estimated to be $\sim$140-180 yr)
inhibits short-term astrometric mass measurements, but the Gliese 337CD system
is ideal for studying the L/T transition at a fixed age and metallicity.
From a compilation of all known widely separated ($\gtrsim$100 AU)
stellar/brown dwarf multiple systems, we find evidence
that the binary fraction of brown dwarfs in these systems
is notably higher
than that of field brown dwarfs, 45$^{+15}_{-13}$\% versus 18$^{+7}_{-4}$\% for analogous samples.
We speculate on possible reasons for this difference, including the
possibility that dynamic (ejection) interactions which may form such wide pairs
preferentially retain binary secondaries due to their greater combined
mass and/or ability to absorb angular momentum.
\end{abstract}


\section{Introduction}

Multiple systems are important laboratories for studying the physical properties
and origins of brown dwarfs, low mass stars incapable of sustaining core hydrogen
fusion \citep{kum62,hay63}.
Brown dwarf binaries facilitate the study of mass or temperature
effects in cool atmospheres independent of age or metallicity variations, and can
be astrometrically and/or spectroscopically monitored to yield system mass measurements
(e.g., Zapatero Osorio et al.\ 2004).
Substellar companions to well-characterized main sequence stars inherit the
age and metallicity of the primary, assuming coevality, providing additional constraints
on mass and composition (e.g., Burgasser et al.\ 2000). Finally,
the multiplicity fraction, separation distribution, and mass ratio distribution of
brown dwarf binaries and brown dwarf companions provide important clues for the
mechanism of their formation (e.g., Close et al.\ 2003).

Brown dwarfs in multiple systems have been found largely in
two distinct populations: as
widely separated ($\rho \gtrsim 20$ AU) companions to stars, forming
low mass ratio systems ($q \equiv {\rm M}_2/{\rm M}_1 \sim$ 0.1); and
as closely separated ($\rho \lesssim 20$ AU) binaries in systems with near-unity
mass ratios.  The paucity of closely separated brown dwarf companions,
most evident in radial velocity monitoring programs, has been termed
the ``Brown Dwarf Desert'' \citep{mrc00}, and a number of mechanisms have been postulated
to explain this trend (e.g., Armitage \& Bonnell 2002).  On the other hand, the
apparent absence of
co-moving widely separated brown dwarf pairs\footnote{\citet{luh04} has
reported the identification of a widely separated
($\rho$ = 240 AU) brown dwarf
pair in the Cha I star-forming region, based on their low-gravity spectra
(implying common cluster membership) and statistically significant
angular proximity.  However, common motion
for these objects has not been verified and the physical association of these objects
remains uncertain.}
has been seen as support
for dynamical ejection models for brown dwarf formation (e.g, Bate, Bonnell, \& Bromm 2002a).

A third group of substellar multiple systems share the properties
of both of these classes: closely separated brown dwarf
binaries that are widely separated
but gravitationally bound to more massive stellar primaries.  Six such
systems are currently known:
Gliese 569Bab \citep{for88,mrt00},
GJ 1001BC \citep{gld99,gol04b},
Gliese 577BC \citep{low01,mcc01,mug04,low05},
Gliese 417BC \citep{kir01a,bou03},
HD 130498BC \citep{pot02,got02},
and $\epsilon$ Ind BC \citep{sch03,mcc03}. All six binaries have high mass ratios
($q \gtrsim 0.6$),
and are separated by at least 45 AU from their apparently single
and more massive main sequence primaries.  One of these systems,
Gliese 569Bab, has been astrometrically and spectroscopically monitored
over a few orbital periods,
providing the first mass measurements for a substellar system \citep{lan01,zap04}; follow-up
observations of HD 130498BC and $\epsilon$ Ind BC are currently underway
(D.\ Potter and M.\ McCaughrean, priv.\ comm.).  By providing multiple,
independent constraints on substellar masses and ages, these companion
binaries provide powerful empirical tests of brown dwarf evolutionary and
structural models \citep{lan01}.

In this article, we present the discovery of a new binary brown dwarf companion system,
Gliese 337CD, identified through high-resolution imaging observations
obtained with the Lick Observatory Adaptive Optics (AO) system.
In $\S$ 2 we describe our observations and data reduction techniques.
In $\S$ 3 we analyze the results, deriving physical properties for the
binary and probable orbital characteristics.  We also
argue for physical association of the brown dwarf pair
based on the proper motion of the Gliese 337 system.
In $\S$ 4 we discuss the wider issue of multiplicity amongst brown dwarf
companions, identifying a significantly higher binary fraction amongst
these sources as compared to field brown dwarfs.
We speculate on how this may be related to the formation of wide binary systems.
Results are summarized in $\S$ 5.

\section{Observations}

The combined system Gliese 337C was originally identified
as the unresolved source 2MASSW J0912145+145940 by \citet{wil01} in the Two Micron
All Sky Survey \citep[hereafter 2MASS]{cut03}.  Optical spectroscopy
indicate a spectral type of L8, sufficiently late to deduce that this
source is substellar.
Gliese 337C is a widely separated ($\rho$ = 43$\arcsec$ $\sim$ 880 AU)
common proper motion companion to the Gliese 337AB system, a G8V+K1V \citep{mas96,ric00,bar00},
nearly equal mass \citep{pou00},
double-lined spectroscopic and visual binary with an orbital separation of 2.4 AU and a period of
2.7 yr \citep{mas96}.  The system lies at a distance of 20.5$\pm$0.4 pc from the Sun
\citep{esa97}. \citet{wil01} estimate an of age 0.6--3.4 Gyr
for Gliese 337AB based on X-ray luminosity and kinematics.  The absolute
$JHK_s$ magnitudes of Gliese 337C are 0.4-0.6 mag brighter than
those of several L8 dwarfs with parallax measurements \citep{dah02,vrb04},
and this source is 0.8 mag brighter at $J$-band than the L8 companion brown
dwarf Gliese 584C
\citep{kir00}.  These measurements have suggested that
Gliese 337C could be an unresolved multiple system.

We observed Gliese 337C as part of a backup natural guide star program
during a single night run with the Lick Observatory AO system on 9 February 2004 (UT).
A log of observations is given in Table 1.
The Lick/LLNL AO system on the 3m Shane Telescope
uses the IRCAL near-infrared imager behind a 127-actuator (61 actively controlled)
deformable mirror.  Corrections to the mirror surface are made through observation
of a V $\lesssim$ 12-13 guide star within 55$\arcsec$ of the target
source onto a Shack-Hartmann wavefront sensor and fast-read CCD camera.
A photo diode quad cell provides tip-tilt corrections.
At reasonable strehl ratios, diffraction-limited ($\sim$0$\farcs$12 at $K_s$)
cores of point sources can be marginally resolved with the 0$\farcs$076 pixel resolution
of IRCAL.  The imaging field of view is 19$\farcs$4 on a side.  Additional information
on the Lick AO system is given in \citet{bau99,bau02} and \citet{gav00}.

Conditions during the night were
poor, with seeing of $\sim 1\arcsec$ at $K$-band and intermittent clouds and cirrus.
We used the bright Gliese 337AB pair (combined $V$ = 6.78) as our tip-tilt and AO
corrector source; its binarity did not affect the AO correction
due to the 2$\arcsec$ pixel scale of the wavefront sensor.
Five 120-second dithered exposures of Gliese 337C were obtained at $K_s$ and three 300-second
dithered exposures were obtained in the narrowband $K_{2.2}$
filter centered at 2.2 $\micron$ (${\Delta}{\lambda}$ = 0.02 $\micron$).
Because of the poor and variable seeing (r$_0$ $\approx$ 5-15 cm) and wide separation between the AO
corrector and science target, strehl ratios were low over the course
of the observations ($\sim$0.06).
The point spread function (PSF) of the observations was
measured by imaging the
single source USNO-B1.0 0992-0177063 \citep{mon03} in both
filters while guiding on the $R$ = 8.1 guide star USNO-B1.0 0992-0177081.
This pair has a similar
separation and position angle ($\rho$ = 45$\farcs$3 at $\theta$ = 276$\degr$)
as that between Gliese 337AB and Gliese 337C ($\rho$ = 46$\farcs$0 at $\theta$ = 262$\degr$).
The PSF full width at half maximum was 0$\farcs$45.

Imaging data was reduced by first constructing dark-subtracted,
median-combined and normalized
flat-field images for both filters
from observations of the twilight sky at the start of the evening.
The sky flat and dark frames were also used to identify dead and hot pixels for
the construction of an pixel mask.  Observations
of the science targets were pair-wise subtracted (using sequential imaging
pairs) to eliminate sky background, then divided by the flat field frame
and corrected for bad pixels by linear interpolation.  Dithered pairs
were combined by shifting (in integer pixel units) and adding,
centering on the peak of each source as verified by
visual inspection.

Reduced images for Gliese 337C and its associated PSF star
in both $K_s$ and $K_{2.2}$ filters are shown in Figure 1.
The extended nature of Gliese 337C is clearly evident along
an ESE/WNW axis.  The PSFs of the components are heavily
blended, however, due to the poor AO correction during the observations.
These images were deconvolved using the PSF fitting code Xphot,
kindly provided by D.\ Koerner.
Starting from the observed PSF of the single star and estimates of the centers of the
components, this code compares model images to the binary data by
iteratively shifting the relative positions and peak fluxes
of the two components.  Convergence occurs when the standard deviation of the difference
image between the data and model image is minimized.  Multiple fits to the $x$- and $y$- pixel
separations and flux ratio between the components were obtained
by using each of the individual PSF
observations (reduced dithered pairs) and varying the initial starting conditions.  The scatter
in these fits was used as an estimate of the measurement uncertainty of the mean value.

For astrometric calibration, we observed the visual double HD 56988AB, selected from
\citet{sha99}.  This pair has a known separation ($\rho$ = 3$\farcs$992$\pm$0$\farcs$004)
and position angle ($\theta$ = 350$\fdg$83$\pm$0$\fdg$10) ideally suited for observation
with IRCAL.  Both components of this double
were bright enough to serve as AO corrector stars ($V \approx 9.0$ for each component),
and we used the southernmost source
for this purpose.  Five dithered exposures were obtained each at $K_s$ and $K_{2.2}$.
Images were reduced as described above.
We used Xphot to measure the separation of the pair, again employing
multiple PSF (in this case, each component of the double)
and initial starting conditions to estimate the empirical uncertainty.
From these measurements, we derived the image pixel scale (0$\farcs$076$\pm$0$\farcs$005)
and orientation (1$\fdg$35$\pm$0$\fdg$15 east of north) of the camera.

\section{Analysis}

\subsection{Common Proper Motion}

While the PSF analysis clearly indicates the presence of two sources
at the position of Gliese 337C, are these sources physical associated?
The Gliese 337 system has a proper motion $\mu$ = 0$\farcs$5789$\pm$0$\farcs$0009 yr$^{-1}$
\citep{esa97}, implying that Gliese 337C itself has moved 3$\farcs$6 between the
time of its detection by 2MASS (18 November 1997 UT) and our observations.
If the second component identified in our observations was an unassociated, stationary
background source, it would have been readily detected as a separate point source by 2MASS.
We performed overlapping PSF simulations on the 2MASS atlas images of Gliese 337C
to determine that an equal magnitude pair can be resolved in those data for
separations $\sim$1$\farcs$5 and greater\footnote{2MASS point source extraction generally resolves
sources only down to 5$\arcsec$ due to blending
(Cutri et al.\ 2003, $\S$I.6.b.ix).}. This limit allows us to
constrain the motion of the second component to 0$\farcs$33-0$\farcs$81 yr$^{-1}$ and
at a position angle within $\pm$25$\degr$ of that of Gliese 337C.
These constraints rule out an extragalactic source for the second component
and strongly suggest common proper motion.

Additionally, there are no optical sources coincident with or moving
toward the current Gliese 337C position in the First or Second Palomar Sky Survey (POSS)
plates \citep{abe59,rei91}.
The second component must therefore have very red optical/near-infrared
colors ($R-K_s \gtrsim 5$), consistent with a late-type star or brown dwarf.
These constraints on the motion and color of the
resolved double, along with their angular proximity,
lead us to conclude that they comprise a gravitationally bound pair.  We hereafter
refer to the combined system as Gliese 337CD.

\subsection{Properties of Gliese 337CD}

Table 2 lists the derived separation, position angle, and $K_s$ and $K_{2.2}$ flux ratios
for Gliese 337CD.
Derived flux ratios were 0.93$\pm$0.10 at $K_s$ and 0.90$\pm$0.08 at $K_{2.2}$,
with the WNW component being slightly fainter in the latter filter.  However, both
measurements are consistent with equal brightness at $K$-band, implying
that Gliese 337CD is likely to be a near-equal mass binary, $q \sim 1$, similar to
the AB components.
Assuming equal brightness across the near infrared,
the individual absolute 2MASS magnitudes of Gliese 337C and D are
$M_J$ = 14.89$\pm$0.09, $M_H$ = 13.78$\pm$0.09 and $M_{K_s}$ = 13.21$\pm$0.07.
These values are consistent with those of Gliese 584C and
the field object 2MASS 1632+1904, both classified in the optical as L8 dwarfs
\citep{kir99,dah02,vrb04}.  However, \citet{vrb04} have shown that $M_{K}$
is effectively constant ($\sim$13.5$\pm$0.5) from L8 to T4, implying
that the secondary could in fact be an early or mid-type T dwarf.  Resolved
imaging and/or spectroscopy can test this possibility.

The astrometric calibrations yield consistent
separations of 0$\farcs$53$\pm$0$\farcs$03 at 292$\pm$7$\degr$ in the $K_s$ images and
0$\farcs$52$\pm$0$\farcs$03 at 290$\pm$8$\degr$ at $K_{2.2}$. The uncertainty is dominated by scatter
amongst the PSF fits.  These values give a projected separation $\rho$ = 10.9$\pm$0.6 AU,
on the high end of the separation distribution amongst known brown dwarf binaries
\citep{bou03,me03b,giz03} and implying a long orbital period.
Using the same mass estimates
from \citet{wil01} of 0.04 $\lesssim$ M $\lesssim$ 0.07 M$_{\sun}$
for each component,
and assuming a semi-major axis $a \approx 1.26\rho$ \citep{fis92},
we estimate an orbital period of 140-180 yr.
Hence, this system is not an ideal target for astrometric monitoring,
although it is amenable for resolved photometric
and spectroscopic investigations.

Assuming reasonable
values for the orbital eccentricity of the CD pair ($\epsilon \lesssim 0.5$),
the Gliese 337ABCD system is dynamically stable as long as
the AB-CD orbital eccentricity is less than 0.8-0.9 \citep{egg95}.
While the wide separation between the binaries suggests that a
highly elliptical orbit is possible,
it is more likely that this system is a long-lived dynamical arrangement.

\section{Discussion}

\subsection{The Binary Fraction of Wide Brown Dwarf Companions}

Gliese 337CD joins a growing list of binary brown dwarf systems that are
widely separated, common motion companions to stellar primaries.
To place this system in context, we compare its properties to those of other known,
widely separated, common motion brown dwarf companions, single or binary (Table 3).
We adopt a lower separation limit of 100 AU for this list, similar to
limits used for most searches of
widely separated systems (e.g., Hinz et al.\ 2002).  This constraint excludes the
companion doubles Gliese 569Bab and
HD 130498BC, a handful of unresolved brown dwarf companions, and planets/brown
dwarfs identified through radial velocity techniques.  Nevertheless,
the seventeen brown dwarf systems listed in Table 3
comprise a useful sample for exploring
the properties of wide brown dwarf companions to nearby stars.

Several characteristics of the
substellar companion binaries are similar
to those of substellar field binaries.  Both groups exhibit
short projected separations ($\rho < 20$ AU)
and near-unity component mass ratios.  The distribution of
separations is also similar, peaking around 2-4 AU
\citep{bou03,giz03} for both groups.  However, the fraction of binaries (${\epsilon}_b \equiv N_{binary}/N_{total}$) is
higher amongst the companions.  For all sources listed in Table 3,
${\epsilon}_b$ = 29$^{+13}_{-8}$\% (5 binaries/17 total\footnote{Statistical uncertainties
for ${\epsilon}_b$ are computed following the prescription of \citet{me03b}.}),
somewhat larger than the binary fraction
typical for field samples, ${\epsilon}_b \approx 10-20$\%  \citep{rei01,bou03,me03b,clo03,giz03}.
Restricting both samples to only brown dwarfs (spectral types L4 and later amongst
the field objects; Gizis et al.\ 2000) with HST or AO observations, the binary fraction gap widens:
${\epsilon}_b = 45^{+15}_{-13}$\% (5/11) for the companion brown dwarfs versus 18$^{+7}_{-4}$\% (9/50)
for the field brown dwarfs, a 2$\sigma$ deviation.
An important caveat to this result is
the fact that neither sample is a complete, unbiased or volume-limited one.
However, as the sources in both samples were generally identified by similar
techniques (compiled from magnitude-limited, color-selected surveys, followed by
high resolution imaging) and have similar distances and separation distributions,
any selection biases are likely to be common.
The dissimilarity in the binary fractions between companion and field brown dwarfs,
at least in the parameter space explored by high resolution imaging, is therefore compelling.

What could account for the higher binary fraction amongst the wide companions?
One current, albeit controversial, model for the formation of brown dwarfs
proposes that early mass accumulation by protostars can be aborted by
dynamic encounters with other prestellar cores and disks, causing
ejection out of the nascent gas cloud \citep{rpt01,bat02}.
In such encounters, the greater the difference in mass between the scattering sources, the more
likely it is that the less massive object (in this case a brown dwarf or brown
dwarf progenitor) will be scattered away from the more massive object.
On the other hand, if the deflected object is a tight brown dwarf binary
system, the larger combined mass of this pair could allow it to be gravitationally captured.
This capture may be facilitated by the transfer of angular momentum into the
brown dwarf binary orbit during the encounter.
Presuming that momentum transfer does not
disrupt the binary pair, such encounters may produce
weakly-bound, widely separated multiple systems more readily
than those involving single brown dwarfs,
leading to a binary excess among the wide companions.
Observable consequences of this process
may include wider and/or more elliptical orbits for
widely separated companion brown dwarf binaries, which may be
tested as further examples of these systems are identified.

The possibility of a higher binary fraction for brown dwarf companions
resonates with the observed enhanced frequency of spectroscopic binaries
among components of visual multiples \citep{tok02}.
The presence of a third star is believed to be integral
to the formation of such close pairs, as
it removes angular momentum
from the binary upon scattering \citep{tok97,bat02b}.
This interaction is somewhat orthogonal to the
mechanism described above, where the low-mass binary
gains angular momentum to remain bound in the wider system.  However,
the similarity of these processes suggests that both could potentially
occur in the high stellar density environment of a star-forming
region.  Simulations exploring these interactions in detail
are necessary to determine their feasibility.
Regardless, it is interesting to speculate
whether the Gliese 337ABCD system
arose through the exchange of angular momentum
between both of its double components.

\subsection{Gliese 337CD and the L/T Transition}

While Gliese 337CD is too wide to
adequately measure its full orbital characteristics in a reasonable period,
this binary does provide
a unique opportunity to study the properties of cool brown dwarf
atmospheres, particularly across the transition between L dwarfs and T dwarfs.
The dramatic shift in near-infrared spectral energy distributions
between late-type L dwarfs -- characterized by continuum dust emission
and H$_2$O and CO bands -- and mid-type T dwarfs -- characterized
by an absence of dust and strong H$_2$O and CH$_4$ bands -- appears
to occur over a fairly narrow effective temperature range \citep{kir00,me02a,gol04a}.
More surprisingly,
early-type T dwarfs are generally brighter than late-type L dwarfs in the 1 $\micron$
region \citep{dah02,tin03,vrb04}, indicating a significant redistribution of
flux at near-constant luminosity.  Cloud condensate models, which adequately
reproduce the overall trends of the L/T transition \citep{ack01,mar02,tsu02,coo03}, cannot
reproduce these seemingly dramatic, apparently temperature-independent, effects.  More elaborate models,
evoking cloud dissipation \citep{me02c} and variations in rainout efficiency
\citep{kna04}, have been proposed but remain controversial \citep{tsu03}.

One complication in addressing this issue is the unknown diversity of physical
properties (mass, age, metallicity) among the current population of
field brown dwarfs.  Differences between these
parameters could potentially blur the fundamental physical process or
processes governing
the L/T transition.  Late-type binaries such as Gliese 337CD,
whose presumably coeval components may straddle the L/T boundary,
are crucial for studying this transition independent of differences
between age or metallicity.  Gliese 337CD is a
particularly vital case,
as the physical parameters of the substellar components
can be constrained by the properties of the stellar primaries.
Resolved photometric and spectroscopic observations of this system
may therefore help explain the changes that occur in
brown dwarf atmospheres as they evolve from dusty L dwarfs to dust-free T dwarfs.

\section{Summary}

We have resolved the companion L8 brown dwarf Gliese 337C using the Lick/LLNL
AO system into a nearly equal-magnitude, closely separated
pair of cool brown dwarfs, hereafter named Gliese 337CD.
The absence of a background source
at the current position of Gliese 337CD either in the original
2MASS and earlier POSS surveys implies common proper motion.  The
long (140-180 yr) orbit of this system is poorly suited for
dynamical mass measurements, but its late spectral type makes it
an important target for studying the L/T transition.  Gliese 337CD
joins a growing list of brown dwarf binaries that are widely
separated but gravitationally bound to more massive stars,
and we find evidence that the frequency of such binaries is higher than that seen
among field brown dwarfs.  Further study of these systems
is warranted, as their orbital, physical and atmospheric characteristics
provide important clues on the formation and evolution
of brown dwarfs and their cool atmospheres.

\acknowledgments

The authors would like to thank telescope operator K.\ Chloros and AO operator E.\ Gates
for their expert assistance during the observations, and the UCO TAC for its generous
allocation of time.
AJB also thanks D.\ Koerner for help with the Xphot program, and
acknowledges useful discussions with J.\ Gizis during the preparation of the manuscript.
Support for this work was provided by NASA through the Spitzer Fellowship Program.
This publication makes use of data from the Two
Micron All Sky Survey, which is a joint project of the University
of Massachusetts and the Infrared Processing and Analysis Center, and
funded by the National Aeronautics and Space Administration and
the National Science Foundation.
2MASS data were obtained from
the NASA/IPAC Infrared Science Archive, which is operated by the
Jet Propulsion Laboratory, California Institute of Technology,
under contract with the National Aeronautics and Space
Administration.

Facilities: \facility{Lick(IRCal,AO)}

\clearpage

\begin{figure}
\epsscale{0.5}
\plotone{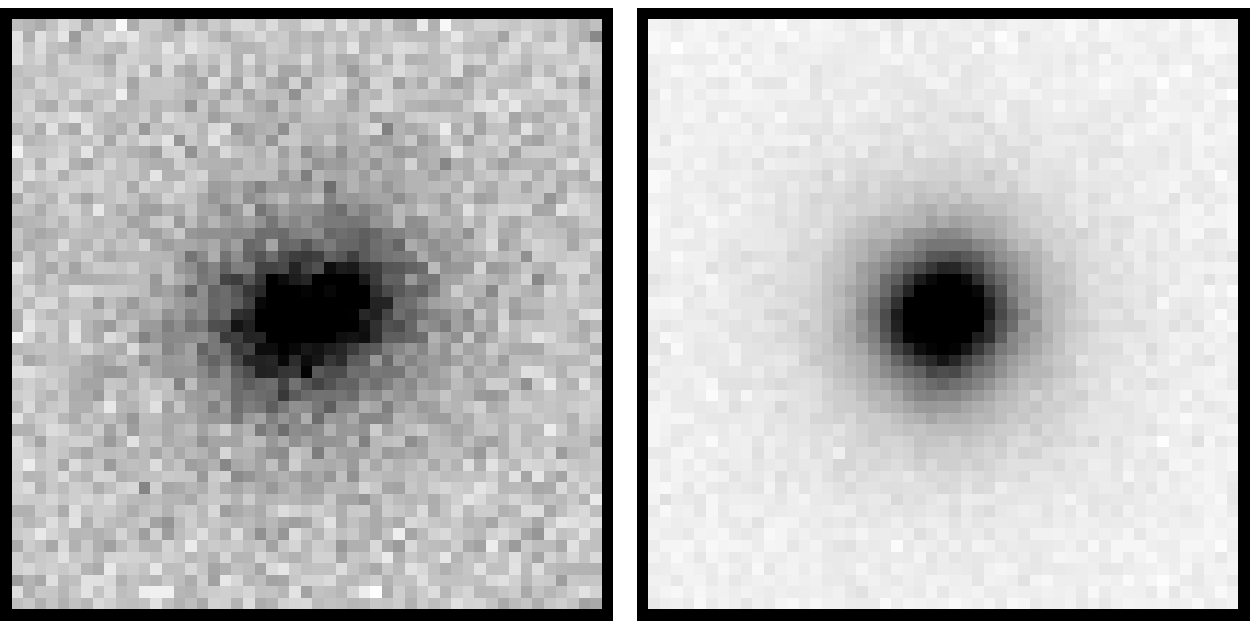}
\plotone{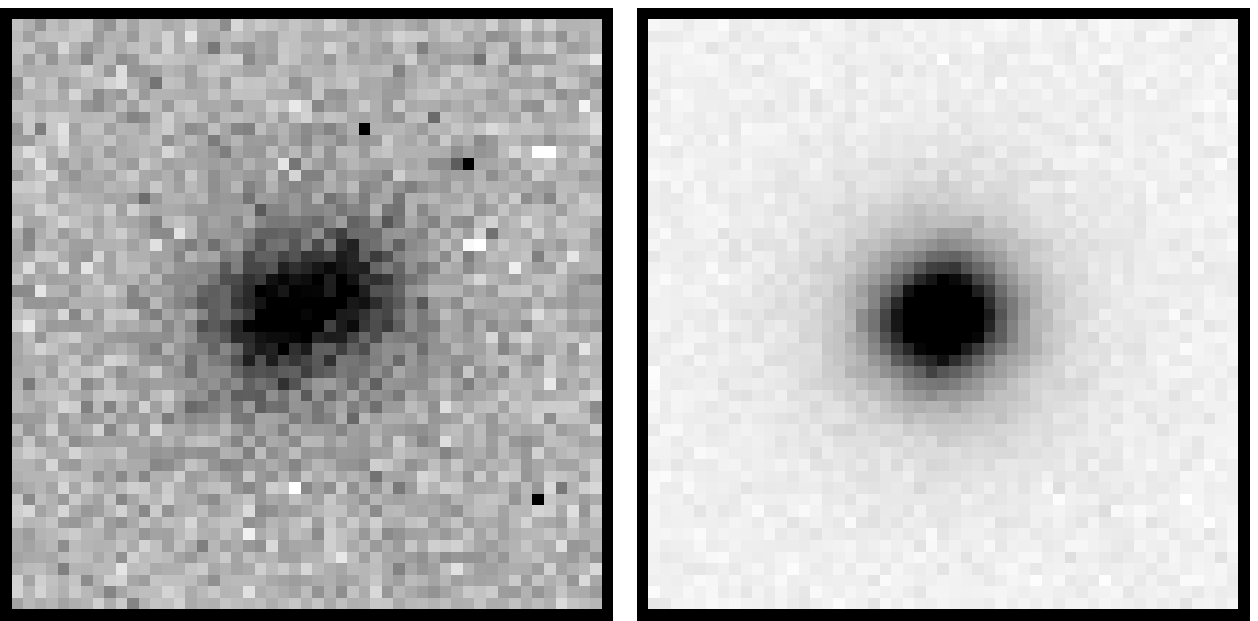}
\caption{IRCal images of Gliese 337CD (left) and the PSF star USNO-B1.0 0992-0177063
(right) in the $K_s$ (top) and $K_{2.2}$ (bottom) filters.  Images are 3$\farcs$9 on a side
and are oriented with north up and east to the left.  The two components of Gliese 337CD
are marginally resolved at 0$\farcs$53$\pm$0$\farcs$03 separation, with the WNW component
being slightly fainter at $K_{2.2}$.
\label{fig1}}
\end{figure}


\begin{deluxetable}{lccclccccccc}
\tabletypesize{\tiny}
\tablecaption{Log of AO Observations on 9 February 2004 (UT).}
\tablewidth{0pt}
\tablehead{
\colhead{Target} &
\colhead{$\alpha$ (J2000)} &
\colhead{$\delta$ (J2000)} &
\colhead{$K_s$\tablenotemark{a}} &
\colhead{AO Corrector} &
\colhead{$\Delta$$\alpha$\tablenotemark{b}} &
\colhead{$\Delta$$\delta$\tablenotemark{b}} &
\colhead{$R$\tablenotemark{c}} &
\colhead{UT} &
\colhead{Filter} &
\colhead{$t$ (s)} &
\colhead{Airmass} \\
\colhead{(1)} &
\colhead{(2)} &
\colhead{(3)} &
\colhead{(4)} &
\colhead{(5)} &
\colhead{(6)} &
\colhead{(7)} &
\colhead{(8)} &
\colhead{(9)} &
\colhead{(10)} &
\colhead{(11)} \\
}
\startdata
HD 56988AB & 07$^h$19$^m$49$\fs$31 & +13$\degr$22$\arcmin$23$\farcs$0 & 8.81$\pm$0.02(A) & HD 56988B & ... & ... & 9.2(A) & 4:19 & $K_s$ & 25 & 1.21 \\
 &  &  &  &  &  &  &  & 4:27 & $K_{2.2}$ & 150 & 1.18  \\
Gliese 337CD & 09$^h$12$^m$14$\fs$69 & +14$\degr$59$\arcmin$39$\farcs$6 & 14.04$\pm$0.06 & Gliese 337AB & -42$\farcs$6 & -5$\farcs$8 & 6.0 & 6:05 & $K_s$ & 600 & 1.21 \\
 &  &  &  &  &  &  &  & 6:20 & $K_{2.2}$ & 900 & 1.18  \\
USNO-B1.0 0992-0177063 & 08$^h$38$^m$26$\fs$49 & +09$\degr$16$\arcmin$58$\farcs$5 & 10.18$\pm$0.02 & USNO-B1.0 0992-0177081 & -44$\farcs$4 & 4$\farcs$8 & 8.1 & 6:41 & $K_{2.2}$ & 240 & 1.16 \\
 &  &  &  &  &  &  &  & 6:47 & $K_s$ & 100 & 1.15  \\
\enddata
\tablenotetext{a}{From 2MASS \citep{cut03}.}
\tablenotetext{b}{Offset from AO corrector star to science target in arcseconds.}
\tablenotetext{c}{From USNO B1.0 \citep{mon03}.}
\end{deluxetable}

\begin{deluxetable}{lll}
\tabletypesize{\small}
\tablecaption{Properties of Gliese 337CD.}
\tablewidth{0pt}
\tablehead{
\colhead{Parameter} &
\colhead{Value} &
\colhead{Ref} \\
}
\startdata
SpT\tablenotemark{a} & L8 & 1 \\
d & 20.5$\pm$0.4 pc & 2 \\
$\mu$ & 0$\farcs$5789$\pm$0$\farcs$0009 yr$^{-1}$ & 2 \\
Age & 0.6-3.4 Gyr & 1 \\
${\rho}_{AB-CD}$ & 43$\arcsec$ & 1 \\
 & 880 AU & 1 \\
${\rho}_{C-D}$ & 0$\farcs$53$\pm$0$\farcs$03 & 3 \\
 & 10.9$\pm$0.6 AU & 3 \\
${\theta}_{C-D}$ & $291{\degr}{\pm}8{\degr}$ & 3 \\
${\Delta}K_s$ & 0.08$\pm$0.12 mag & 3 \\
${\Delta}K_{2.2}$ & 0.11$\pm$0.10 mag & 3 \\
Period & $\sim$140-180 yr & 3 \\
\enddata
\tablenotetext{a}{Optical spectral type from \citet{wil01}.}
\tablerefs{(1) \citet{wil01}; (2) \citet{esa97}; (3) This paper.}
\end{deluxetable}

\begin{deluxetable}{lllccccccl}
\tabletypesize{\scriptsize}
\tablecaption{Widely Separated Brown Dwarf Companion Systems\tablenotemark{a}.}
\tablewidth{0pt}
\tablehead{
\colhead{Name} &
\multicolumn{2}{c}{Spectral Types} &
\multicolumn{2}{c}{${\rho}_{*-BD}$} &
\multicolumn{2}{c}{${\rho}_{BD-BD}$\tablenotemark{b}} &
\colhead{$q$} &
\colhead{Age} &
\colhead{Ref} \\
\colhead{} &
\colhead{Primaries} &
\colhead{Secondaries} &
\colhead{($\arcsec$)} &
\colhead{(AU)} &
\colhead{($\arcsec$)} &
\colhead{(AU)} &
\colhead{} &
\colhead{(Gyr)} &
\colhead{} \\
\colhead{(1)} &
\colhead{(2)} &
\colhead{(3)} &
\colhead{(4)} &
\colhead{(5)} &
\colhead{(6)} &
\colhead{(7)} &
\colhead{(8)} &
\colhead{(9)} &
\colhead{(10)}  \\
}
\startdata
TWA 5B & M1.5V & M8.5V & 2.0 & 100 & $\lesssim$ 0.15 & $\lesssim$ 8 & ... & 0.02-0.3 & 1,2 \\
GD 165B & DA4 & L4 & 3.7 & 120 & $\lesssim$ 1 & $\lesssim$ 30 & ... & 1.2-5.5 & 3,4,5 \\
GJ 1001BC & M4V & L5+L5\tablenotemark{c} & 18 & 170 & 0.09 & 0.8 & $\sim$1\tablenotemark{c} & 1-10 &  6,7 \\
* $\eta$ Tel B\tablenotemark{d} & A0V & M7.5V & 4.2 & 190 & $\lesssim$ 0.15 & $\lesssim$ 7 &  ... & 0.02-0.03 & 8,9 \\
GSC 08047-00232B & K3V & M6-9V: & 3.2 & 200 & $\lesssim$ 0.1 & $\lesssim$ 7 & ... & 0.01-0.04 & 10,11,12 \\
GG Tau Bb\tablenotemark{e} & M5V & M7V & 1.5 & 210 & $\lesssim$ 0.1 & $\lesssim$ 14 & ... & 0.01-0.02 & 13,14 \\
Gliese 577BC & G5V & M5V:+M5V:\tablenotemark{c,}\tablenotemark{f} & 5.4 & 240 & 0.08 & 3.6 & $\sim$1\tablenotemark{c} & 0.3-0.5 & 15,16,17,18 \\
GJ 1048B & K2V & L1 & 12 & 250 & $\lesssim$ 1 & $\lesssim$ 20 & ... & 0.6-1.0 & 19,20 \\
G 196-3B & M2.5V & L2 & 16 & 320 & $\lesssim$ 1 & $\lesssim$ 20 & ... & 0.02-0.3 & 21 \\
GJ 4287B\tablenotemark{g} & M0V+M0V\tablenotemark{c} & M9.5V & 34 & 640 & $\lesssim$ 0.3 & $\lesssim$ 6 & ... & 0.6-10 & 22,23 \\
LP 261-75B\tablenotemark{h} & M: & L6 & 13 & 820 & $\lesssim$ 1 & $\lesssim$ 60 & ... & 1-10 & 24,25,26 \\
{\bf Gliese 337CD} & G8V+K1V & L8+L8/T:\tablenotemark{c} & 43 & 880 & 0.53 & 10.9 & $\sim$1\tablenotemark{c} & 0.6-3.4 & 26,27 \\
$\epsilon$ Ind BC & K5V & T1+T6 & 402 & 1460 & 0.62 & 2.2 & $\sim$0.6  & 0.8-2.0  & 28,29 \\
Gliese 570D & K4V+M1.5V+M3V & T8 & 258 & 1530 & $\lesssim$ 0.1 & $\lesssim$ 0.6 & ... & 2-5 & 30,31 \\
Gliese 417BC & G0V & L4.5+L6:\tablenotemark{c} & 90 & 2000 & 0.07 & 1.5 & $\sim$0.7\tablenotemark{c} & 0.08-0.3 & 32,33 \\
HD 89744B & F7V/IV & L0 & 63 & 2460 & $\lesssim$ 1 & $\lesssim$ 40 & ... & 1.5-3.0 & 27,33 \\
Gliese 584C & G1V+G3V & L8 & 194 & 3600 & $\lesssim$ 1 & $\lesssim$ 20 & ... & 1.0-2.5 & 32 \\
\enddata
\tablenotetext{a}{Defined here as common proper or radial motion
systems with star-brown dwarf separations greater than 100 AU.}
\tablenotetext{b}{Upper limits on the apparent separations for equal-magnitude unresolved sources are taken from
the literature or assumed to be 1$\arcsec$ in the case of no high resolution imaging follow-up.}
\tablenotetext{c}{Estimated from differential photometry.}
\tablenotetext{d}{Also known as HR 7329B \citep{low00}.}
\tablenotetext{e}{Also known as GG Tau/c, this pair is separated by 1400 AU from the binary stellar
pair GG Tau Aab.  All four components share common radial motion
and are believed to have formed
from the same cloud core \citep{whi99}.}
\tablenotetext{f}{\citet{mug04} give a spectral type M4.5 for the combined system Gliese 577BC,
stating that it is a 0.16-0.20 M$_{\sun}$ low mass star.  \citet{low05} determine a spectral
type of M5.5 and component masses of 0.08 M$_{\sun}$.  \citet{mcc01} classify the source as M5 and at
the stellar/substellar boundary.  As such, we include this source as a {\it possible} double brown
dwarf companion system.}
\tablenotetext{g}{Also known as G 216-7B \citep{kir01b}.}
\tablenotetext{h}{Also known as 2MASSW J0951054+355801 \citep{kir00}, this object's proper motion
as measured by \citet{vrb04},
$\mu = 0{\farcs}189{\pm}0{\farcs}011$ yr$^{-1}$ at $\theta = 211{\fdg}8{\pm}1{\fdg}6$,
is consistent with that of LP 261-75 from the revised NLTT catalog \citep{sal03},
$\mu = 0{\farcs}203{\pm}0{\farcs}013$ yr$^{-1}$ at $\theta = 209{\fdg}9{\pm}1{\fdg}5$.
Given its close angular proximity (13$\farcs$3) and common motion, these objects are likely
to be gravitationally bound.  The measured distance of LP 261-75B, d = 60$\pm$30 pc \citep{vrb04},
places a poor constraint on the projected separation of this low-mass stellar-brown dwarf
pair, but it is clearly a widely separated system.}
\tablerefs{(1) \citet{low99}; (2) \citet{neh00}; (3) \citet{bec88}; (4) \citet{kir93}; (5) \citet{kir99b};
(6) \citet{gld99}; (7) \citet{gol04b}; (8) \citet{low00}; (9) \citet{gue01};
(10) \citet{cha03}; (11) \citet{neh03b}; (12) \citet{neh04};
(13) \citet{lei91}; (14) \citet{whi99};
(15) \citet{low01}; (16) \citet{mcc01}; (17) \citet{mug04};
(18) \citet{low05}; (19) \citet{giz01}, (20) \citet{sei04};
(21) \citet{reb98}; (22) \citet{kir01b}; (23) Burgasser \& Kirkpatrick, in prep.; (24) \citet{kir00};
(25) \citet{vrb04}; (26) This paper; (27) \citet{wil01};
(28) \citet{sch03}; (29) \citet{mcc03}; (30) \citet{me00a}; (31) \citet{me03b};
(32) \citet{kir01a}; (33) \citet{bou03} }
\end{deluxetable}


\begin{thebibliography}{}

\bibitem[Abell(1959)]{abe59}Abell, G.\ O. 1959, \pasp, 67, 258

\bibitem[Ackerman \& Marley(2001)]{ack01}Ackerman, A.\ S., \& Marley, M.\ S.
2001, \apj, 556, 872

\bibitem[Armitage \& Bonnell(2002)]{arm02}Armitage, P.\ J., \& Bonnell, I.\ A.
2002, \mnras, 330, L11

\bibitem[Barnaby et al.(2000)]{bar00} Barnaby, D., Spillar, E., Christou, J. C.,
\& Drummond, J. D. 2000, \aj, 119, 378

\bibitem[Bate, Bonnell, \& Bromm(2002a)]{bat02} Bate, M.\ R., Bonnell, I.\ A.,
\& Bromm, V. 2002a, \mnras, 332, L65

\bibitem[Bate, Bonnell, \& Bromm(2002b)]{bat02b} ---. 2002b \mnras, 336, 705

\bibitem[Bauman et al.(1999)]{bau99} Bauman, B.\ J., Gavel, D.\ T., Waltjen, K.\ E.,
Freeze, G.\ J., Keahi, K.\ A., Kuklo, T.\ C., Lopes, S.\ K., Newman, M.\ J., \& Olivier, S.\ S.
1999, SPIE, 3762, 194

\bibitem[Bauman et al.(2002)]{bau02} Bauman, B.\ J., Gavel, D.\ T., Waltjen, K.\ E.,
Freeze, G.\ J., Hurd, R.\ L., Gates, E.\ L., Max, C.\ E., Olivier, S.\ S., \& Pennington, D.\ M.
2002, SPIE, 4494, 19

\bibitem[Becklin \& Zuckerman(1988)]{bec88}Becklin, E.\ E., \& Zuckerman,
B. 1988, Nature, 336, 656

\bibitem[Bouy et al.(2003)]{bou03} Bouy, H., Brandner, W., Mart{\'{\i}}n, E.\ L.,
Delfosse, X., Allard, F., \& Basri, G. 2003, \aj, 126, 1526

\bibitem[Burgasser et al.(2003)]{me03b}Burgasser, A.\ J., Kirkpatrick,
J.\ D., Reid, I.\ N., Brown, M.\ E., Miskey, C.\ L., \& Gizis, J.\ E. 2003b,
\apj, 586, 512

\bibitem[Burgasser et al.(2002a)]{me02c}Burgasser, A.\ J., Marley, M.\ S.,
Ackerman, A.\ S., Saumon, D.,
Lodders, K., Dahn, C.\ C., Harris, H.\ C., \& Kirkpatrick, J.\ D.
2002b, \apj, 571, L151

\bibitem[Burgasser et al.(2000)]{me00a} Burgasser, A.\ J., et al. 2000, \apj, 531, L57

\bibitem[Burgasser et al.(2002b)]{me02a} ---. 2002b, \apj, 564, 421

\bibitem[Chauvin et al.(2003)]{cha03}Chauvin, G., Thomson, M., Dumas, C., Beuzit, J.-L., Lowrance, P.,
Fusco, T., Lagrange, A.-M., Zuckerman, B., \& Mouillet, D. 2003, \aap, 404, 157

\bibitem[Close et al.(2003)]{clo03}Close, L.\ M., Siegler, N., Freed, M.,
\& Biller, B. 2003, \apj, 587, 407

\bibitem[Cooper et al.(2003)]{coo03}Cooper, C.\ S., Sudarsky, D., Milson, J.\ A.,
Lunine, J.\ I., \& Burrows, A. 2003, \apj, 586, 1320

\bibitem[Cutri et al.(2003)]{cut03}Cutri, R.\ M., et al. 2003, Explanatory Supplement
to the 2MASS All Sky Data Release,
\url{http://www.ipac.caltech.edu/2mass/releases/allsky/doc/explsup.html}

\bibitem[Dahn et al.(2002)]{dah02}Dahn, C.\ C., et al. 2002, \aj,
124, 1170

\bibitem[Eggleton \& Kiseleva(1995)]{egg95}Eggleton, P., \& Kiseleva, L. 1995, \apj,
455, 640

\bibitem[ESA(1997)]{esa97} ESA, 1997, The Hipparcos and Tycho Catalogues, ESA SP-1200

\bibitem[Fischer \& Marcy(1992)]{fis92}Fischer, D.\ A., \& Marcy, G.\ W. 1991, \apj, 396, 178

\bibitem[Forrest, Skrutskie, \& Shure(1988)]{for88}Forrest, W.\ J.,
Skrutskie, M.\ F., \& Shure, M. 1988, \apj, 330, L119

\bibitem[Gavel et al.(2000)]{gav00}Gavel, D.\ T., Olivier, S.\ S.., Bauman, B.\ J.,
Max, C.\ E., \& Macintosh, B. A. 2000, SPIE, 4007, 63


\bibitem[Gizis, Kirkpatrick, \& Wilson(2001)]{giz01}Gizis, J.\ E.,
Kirkpatrick, J.\ D., \& Wilson, J.\ C. 2001, \aj, 121, 2185

\bibitem[Gizis et al.(2000)]{giz00}Gizis, J.\ E., Monet, D.\ G., Reid, I.\ N.,
Kirkpatrick, J.\ D., Liebert, J., \& Williams, R. 2000, \aj, 120, 1085

\bibitem[Gizis et al.(2003)]{giz03}Gizis, J.\ E., Reid, I.\ N., Knapp, G.\ R., Liebert, J.,
Kirkpatrick, J.\ D., Koerner, D.\ W., \& Burgasser, A.\ J. 2003, \aj, 125, 3302

\bibitem[Goldman et al.(1999)]{gld99}Goldman, B., et al. 1999, \aap, 351,
L5

\bibitem[Golimowski et al.(2004a)]{gol04a} Golimowski, D.\ A., et al. 2004, \aj, 127, 3516

\bibitem[Golimowski et al.(2004b)]{gol04b} ---. 2004, \aj, 128, 1733

\bibitem[Goto et al.(2002)]{got02}Goto, M., et al. 2002, 567, L59

\bibitem[Guenther et al.(2001)]{gue01}Guenther, E. W., Neuh{\"{a}}user, R., Hu{\'{e}}lamo, N.,
Brandner, W., \& Alves, J. 2001, \aap, 365, 514

\bibitem[Hayashi \& Nakano(1963)]{hay63}Hayashi, C., \& Nakano, T. 1963, Prog.\
Theo.\ Physics, 30, 4

\bibitem[Hinz et al.(2002)]{hin02}Hinz, J.\ L., McCarthy, D.\ W., Jr., Simons, D.\ A., Henry, T.\ J.,
Kirkpatrick, J.\ D., \& McGuire, P.\ C. 2002, \aj, 123, 2027

\bibitem[Kirkpatrick et al.(2001a)]{kir01a}Kirkpatrick, J.\ D., Dahn, C.\ C.,
Monet, D.\ G., Reid, I.\ N., Gizis, J.\ E.,
Liebert, J., \& Burgasser, A.\ J. 2001a,
\aj, 121, 3235

\bibitem[Kirkpatrick et al.(2001b)]{kir01b}Kirkpatrick, J.\ D., Liebert, J.,
Cruz, K.\ L., Gizis, J.\ E., \& Reid, I.\ N. 2001b, \pasp, 113, 814

\bibitem[Kirkpatrick et al.(2000)]{kir00}Kirkpatrick, J.\ D., Reid, I.\ N.,
Liebert, J., Gizis, J.\ E., Burgasser, A.\ J., Monet, D.\ G., Dahn, C.\ C.,
Nelson, B., \& Williams, R.\ J. 2000, \aj, 120, 447

\bibitem[Kirkpatrick et al.(1999a)]{kir99b}Kirkpatrick, J.\ D., Allard, F.,
Bida, T., Zuckerman, B., Becklin, E.\ E., Chabrier, G., \& Baraffe, I.
1999a, \apj, 519, 834

\bibitem[Kirkpatrick et al.(1993)]{kir93}Kirkpatrick, J.\ D., Kelly, D.\ M.,
Rieke, G.\ H., Liebert, J., Allard, F., \& Wehrse, R. 1993, \apj, 402, 643

\bibitem[Kirkpatrick et al.(1999b)]{kir99}Kirkpatrick, J.\ D., et al. 1999b,
\apj, 519, 802

\bibitem[Knapp et al.(2004)]{kna04}Knapp, G., et al. 2004, \apj, 127, 3553

\bibitem[Kumar(1962)]{kum62}Kumar, S.\ S. 1962, \aj, 67, 579

\bibitem[Lane et al.(2001)]{lan01} Lane, B.\ F., Zapatero Osorio, M.\ R.,
Britton, M.\ C., Mart{\'{\i}}n, E.\ L., \& Kulkarni, S.\ R. 2001, \apj,
560, 390

\bibitem[Leinert et al.(1991)]{lei91}Leinert, Ch., Haas, M., Mundt, R., Richichi, A.,
\& Zinnecker, H. 1991, \aap, 250, 407

\bibitem[Lowrance(2001)]{low01} Lowrance, P.\ J. 2001, Ph.D. Thesis,
University of California Los Angeles

\bibitem[Lowrance et al.(1999)]{low99}Lowrance, P.\ J., et al. 1999,
\apj, 512, L69

\bibitem[Lowrance et al.(2000)]{low00} ---. 2000,
\apj, 541, 390

\bibitem[Lowrance et al.(2005)]{low05} ---. 2005, submitted

\bibitem[Luhman(2004)]{luh04}Luhman, K.\ L. 2004, \apj, 614, 398

\bibitem[Marcy \& Butler(2000)]{mrc00}Marcy, G.\ W., \& Butler, R.\ P. 2000,
\pasp, 112, 137

\bibitem[Marley et al.(2002)]{mar02}Marley, M.\ S., Seager, S., Saumon, D., Lodders, K.,
Ackerman, A.\ S., Freedman, R., \& Fan, X. 2002, \apj, 568, 335

\bibitem[Mart{\'{\i}}n et al.(2000)]{mrt00}Mart{\'{\i}}n, E.\ L., Koresko,
C.\ D., Kulkarni, S.\ R., Lane, B.\ F., \&  Wizinowich, P.\ L. 2000, \apj,
529, L37

\bibitem[Mason, McAllister, \& Hartkopf(1996)]{mas96} Mason, B.\ D., McAllister, H.\ A.,
\& Hartkopf, W.\ I. 1996, \aj, 112, 276

\bibitem[McCarthy, Zuckerman, \& Becklin(2001)]{mcc01}McCarthy, C., Zuckerman, B.,
\& Becklin, E.\ E. 2001, \aj, 121, 3259

\bibitem[McCaughrean et al.(2003)]{mcc03}McCaughrean, M., Close, L.\ M.,
Scholz, R.-D., Lenzen, R., Biller, B., Brandner, W., Hartung, M., \& Lodieu, N.
2003, \aap, 413, 1029

\bibitem[Monet et al.(2003)]{mon03} Monet, D.\ G. 2003, \aj, 125, 984 (USNO-B1.0 Catalog)


\bibitem[Mugrauer et al.(2004)]{mug04}Mugrauer, M., et al. 2004, \aap, 417, 1031



\bibitem[Neuh\"{a}user \& Guenther(2004)]{neh04}Neuh\"{a}user, R. \& Guenther, E.\ W.
2004, \aap, 440, 627

\bibitem[Neuh\"{a}user et al.(2003)]{neh03b}Neuh\"{a}user, R., Guenther, E. W.,
Alves, J., Hu{\'{e}}lamo, N., Ott, Th., Eckart, A. 2003, AN, 324, 535


\bibitem[Neuh\"{a}user et al.(2000)]{neh00}Neuh\"{a}user, R., Guenther, E.\ W.,
Petr, M.\ G., Brandner, W., Hu\'{e}lamo, N., \& Alves, J. 2000, \aap, 360, L39

\bibitem[Pourbaix(2000)]{pou00}Pourbaix, D. 2000, \aaps, 145, 215

\bibitem[Potter et al.(2002)]{pot02}Potter, D., Mart{\'{\i}}n, E.\ L.,
Cushing, M.\ C., Baudoz, P., Brandner, W., Guyon, O., \& Neuh\"{a}user, R.
2002, \apj, 567, L133

\bibitem[Rebolo et al.(1998)]{reb98}Rebolo, R., Zapatero Osorio, M.\ R.,
Madruga, S., B{\'{e}}jar, V.\ J.\ S., Arribas, S., \& Licandro, J.
1998, Science, 282, 1309

\bibitem[Reid et al.(2001)]{rei01}Reid, I.\ N., Gizis, J.\ E., Kirkpatrick,
J.\ D., \& Koerner, D. 2001, \aj, 121, 489

\bibitem[Reid et al.(1991)]{rei91}Reid, I.\ N., et al. 1991, \pasp, 103, 661

\bibitem[Reipurth \& Clarke(2001)]{rpt01}Reipurth, B., \& Clarke, C. 2001, \aj, 122, 432

\bibitem[Richichi et al.(2000)]{ric00}Richichi, A., Ragland, S., Calamai, G., Richter, S.,
\& Stecklum, B. 2000, \aap, 361, 594

\bibitem[Salim \& Gould(2003)]{sal03}Salim, S., \& Gould, A. 2003, \apj, 582, 1011

\bibitem[Scholz et al.(2003)]{sch03}Scholz, R.-D., McCaughrean, M.\ J.,
Lodieu, N., \& Kuhlbrodt, B. 2003, \aap, 398, L29

\bibitem[Seifahrt, Neuh{\"{a}}user, \& Mugrauer(2004)]{sei04}Seifahrt, A.,  Neuh{\"{a}}user, R., \& Mugrauer, M.
2004, \aap, 421, 255

\bibitem[Shatsky et al.(1999)]{sha99}Shatsky, N., Sinachopoulos, D., Prado, P., \& van Dessel, E.
1999, \aaps, 139, 69

\bibitem[Tinney, Burgasser, \& Kirkpatrick(2003)]{tin03}Tinney, C.\ G., Burgasser, A.\ J.,
\& Kirkpatrick, J.\ D. 2003, \aj, 126, 975

\bibitem[Tokovinin(1997)]{tok97}Tokovinin, A.\ A. 1997, AstL, 23, 727

\bibitem[Tokovinin \& Smekhov(2002)]{tok02}Tokovinin, A.\ A., \& Smekhov, M.\ G. 2002, \aap, 382, 118

\bibitem[Tsuji(2002)]{tsu02}Tsuji, T. 2002, \apj, 575, 264

\bibitem[Tsuji \& Nakajima(2003)]{tsu03}Tsuji, T., \& Nakajima, T. 2003, \apj, 585, L151

\bibitem[Vrba et al.(2004)]{vrb04}Vrba, F.\ J., et al. 2004, \aj, 127, 2948

\bibitem[White et al.(1999)]{whi99}White, R.\ D., Ghez, A.\ M., Reid, I.\ N.,
\& Schultz, G. 1999, \apj, 520, 811

\bibitem[Wilson et al.(2001)]{wil01}Wilson, J.\ C., et al. 2001, \aj,
122, 1989

\bibitem[Zapatero Osorio et al.(2004)]{zap04}Zapatero Osorio, M.\ R., Lane, B.\ F., Pavlenko, Ya.,
Mart{\'{i}}n, E. L., Britton, M., \& Kulkarni, S. R. 2004, \apj, 615, 958

\end{thebibliography}
\end{document}